\documentclass[aps,prb,twocolumn,superscriptaddress,preprintnumbers,amsmath,amssymb]{revtex4-1}
\usepackage{bm}
\usepackage{graphicx}
\usepackage{natbib}
\draft

\begin{document}

\title{Hall-effect within the colossal magnetoresistive semi-metallic state of
MoTe$_2$}

\author{Qiong Zhou}
\affiliation{National High Magnetic Field Laboratory, Florida
State University, Tallahassee-FL 32310, USA}
\affiliation{Department of Physics, Florida State University, Tallahassee-FL 32306, USA}

\author{D. Rhodes}
\affiliation{National High Magnetic Field Laboratory, Florida
State University, Tallahassee-FL 32310, USA}
\affiliation{Department of Physics, Florida State University, Tallahassee-FL 32306, USA}

\author{Q. R. Zhang}
\affiliation{National High Magnetic Field Laboratory, Florida
State University, Tallahassee-FL 32310, USA}
\affiliation{Department of Physics, Florida State University, Tallahassee-FL 32306, USA}

\author{S. Tang}
\affiliation{National High Magnetic Field Laboratory, Florida
State University, Tallahassee-FL 32310, USA}

\author{R. Sch{\"o}nemann}
\affiliation{National High Magnetic Field Laboratory, Florida
State University, Tallahassee-FL 32310, USA}

\author{L. Balicas}
\email[]{balicas@magnet.fsu.edu}
\affiliation{National High Magnetic Field Laboratory, Florida
State University, Tallahassee-FL 32310, USA}
\date{\today}

\begin{abstract}
Here, we report a systematic study on the Hall-effect of the semi-metallic state of bulk
MoTe$_2$, which was recently claimed to be a candidate for a novel type of Weyl semi-metallic state.
The temperature ($T$) dependence of the carrier densities and of their mobilities, as estimated from
a numerical analysis based on the isotropic two-carrier model, indicates
that its exceedingly large and non-saturating magnetoresistance
may be attributed to a near perfect compensation between the densities of electrons and holes at low temperatures. A sudden increase in hole density, with a concomitant
rapid increase in the electron mobility below $T \sim 40$ K, leads to comparable densities of electrons and holes
at low temperatures suggesting a possible electronic phase-transition around this temperature.
\end{abstract}


\maketitle 

A Weyl semi-metal (WSM)\cite{huang2015weyl,lv2015experimental,huang2015observationTaAs,
xu2015discoveryNbAs,xu2015discoverywely} is a topological semi-metallic state with low energy excitations whose electronic
bands disperse linearly along the three dimensional momentum space from
a node called a Weyl point. Their existence is associated with the lack of time-reversal or inversion symmetry. 
These Weyl points act as topological charges that means as sources and drains of Berry phase. These compounds are often characterized by anomalous transport properties or the existence of surface states \cite{huang2015weyl,lv2015experimental,
xu2015discoveryNbAs,xu2015discoverywely}. For example, they frequently exhibit extremely large magnetoresistance\cite{huang2015observationTaAs} (MR) which surpasses by far the one observed in thin metallic films displaying the giant magnetoresistive effect, or the magnitude of the MR observed in Cr-based chalcogenide spinels or in Mn-based pervoskites\cite{jin1994colossal,ramirez1997colossal,matsumoto2001room}.
In the past, materials displaying large MR attracted a lot of attention due to their potential for applications in engineering or in information technology either as sensitive magnetic sensors or, for instance, as the basic elements in magnetic random access memories\cite{prinz1998magnetoelectronics,parkin2003magnetically}.

WTe$_2$, MoTe$_2$ and their alloys W$_{1-x}$Mo$_x$Te$_2$ were introduced as candidates for a novel type of Weyl semi-metallic state called Weyl type-II \cite{soluyanov2015type,chang2016prediction}. Distinct from the previously discussed type-I WSMs (e.g. TaAs, NbAs, NbP, TaP), which are characterized by
pairs of linearly dispersing touching points between the valence and the conduction bands, the Weyl points in type-II WSMs
result from linear touching points at the boundary between electron and hole pockets.
The transport properties of these type-II WSM candidates have attracted considerable attention due to the observation of
extremely large and non-saturating MR in WTe$_2$ \cite{ali2014large,keum2015bandgap,rhodes2015role} and in MoTe$_2$ \cite{hughes1978electrical,canadell1990semimetallic,zandt2007quadratic,rhodes2016impurity}.
The coexistence of electron and hole pockets of approximately
the same size was observed in WTe$_2$ by angle-resolved
photo-emission spectroscopy (ARPES) \cite{pletikosic2014electronic,jiang2015signature}, suggesting near perfect carrier compensation at low temperatures. Perfect carrier compensation was claimed\cite{ali2014large} to be responsible for the magnetoresistivity of WTe$_2$. This is partially supported by the geometry of the Fermi surface extracted from Shubnikov de Haas (SdH) quantum
oscillations \cite{cai2015drastic}, and by a two-band analysis of the Hall-effect \cite{luo2015hall}. Nevertheless,
the two-carrier analysis of the Hall-effect indicates that the fraction of holes is $\sim 0.9$ times that of electrons,
contradicting the claims of Ref. \onlinecite{ali2014large} of near perfect carrier compensation in WTe$_2$.

Our study focuses on MoTe$_2$, the sister compound of WTe$_2$. MoTe$_2$ is a layered transition
metal dichalcogenide which can crystallize into two different structures:
the $2H$ or the hexagonal $\alpha$-phase \cite{revolinsky1966electrical}, and the $1T^{\prime}$ or the monoclinic $\beta$-phase\cite{revolinsky1966electrical} which
undergoes a structural phase-transition below 240 K into an orthorhombic phase frequently referred to as the $T_{d}$-phase\cite{dawson1987electronic}. Throughout this manuscript we will refer to this low temperature orthorhombic phase of MoTe$_2$ as the $\gamma-$phase or simply as $\gamma-$MoTe$_2$\cite{dawson1987electronic}. The $2H$ phase is semiconducting with the Mo atoms being trigonal-prismatically coordinated by Te atoms forming stacked layers which couple through a weak, or a van der
Waals like interaction. The $\beta$ phase crystallizes in a monoclinic space group while the $\gamma$ phase has the aforementioned orthorhombic structure\cite{clarke1978low}. In contrast, WTe$_2$ is known to crystallize only in an orthorhombic phase with a structure which is akin to that of $\gamma-$MoTe$_2$\cite{dawson1987electronic}.

Currently, $\gamma-$MoTe$_2$ is attracting a lot of interest due to its extremely large magnetoresistivity \cite{keum2015bandgap},
the observation of pressure-driven superconductivity\cite{qi2015superconductivity},
and the aforementioned prediction for a Weyl type-II semi-metallic state\cite{soluyanov2015type,sun2015prediction}.
Recent ARPES measurements \cite{xu2016discovery,liang2016electronic,jiang2016observation,tamai2016fermi,huang2016spectroscopic}
claim to observe Fermi arcs at the surface of $\gamma-$MoTe$_2$ and to uncover a bulk
electronic structure in broad agreement with the band structure calculations, and consequently
with the predicted type-II WSM state in this system. The observation of a non-trivial Berry phase through quantum oscillatory phenomena indicates that its electronic structure is indeed topological although it reveals a Fermi surface topography markedly distinct from the calculated one \cite{rhodes2016impurity}. Furthermore, the observation of superconductivity in $\gamma-$MoTe$_2$ at ambient pressure, with an impurity-dependent transition temperature, points towards
an unconventional and probably topological superconducting state \cite{rhodes2016impurity}.
\begin{figure*}[htb]
\begin{center}
\includegraphics[width= 14 cm]{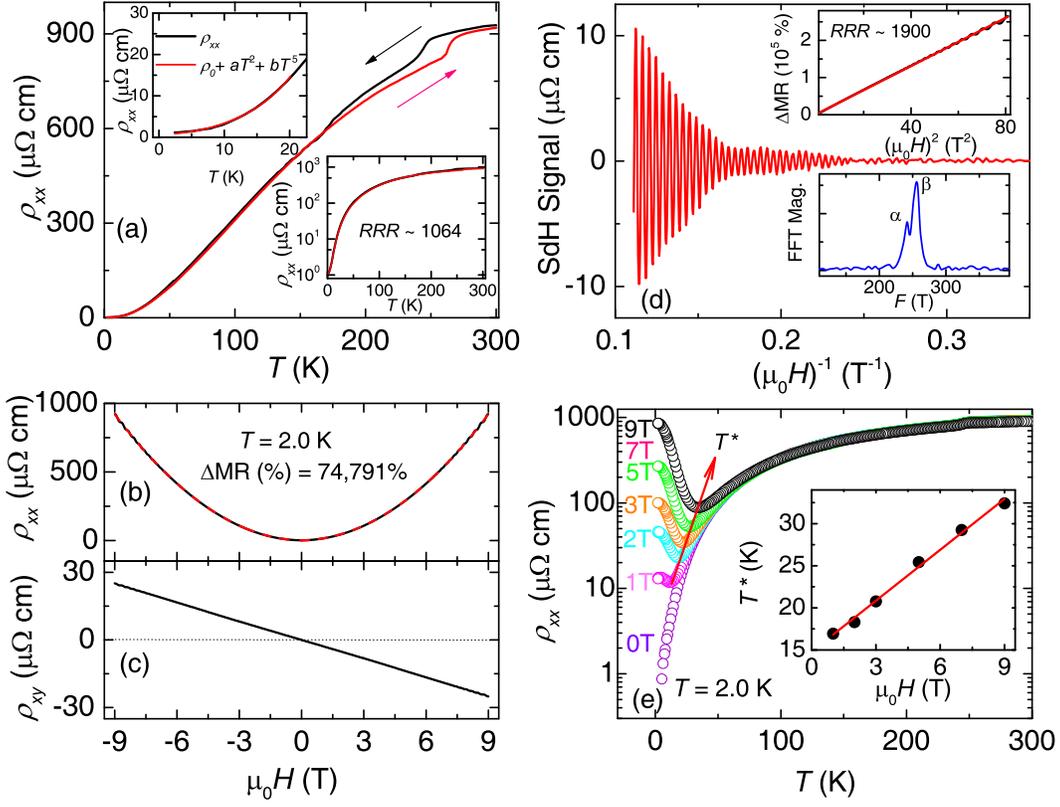}
\caption{(a) Longitudinal resistivity $\rho_{xx}$ of semi-metallic MoTe$_2$ under zero magnetic field and as a function of the temperature. A hysteretic
anomaly is observed around $\sim 240$ K, which is associated with the transition from the monoclinic $1T^{\prime}$
to the orthorhombic $T_{d}$ structure. Upper inset: fit of $\rho_{xx}(T)$ to a combination of Fermi liquid and electron-phonon scattering mechanisms.
Lower inset: $\rho_{xx}(T)$ in a logarithmic scale indicating a resistivity ratio $RRR$ = 1064.
(b) and (c) $\rho_{xx}$ and $\rho_{xy}$ as functions of $\mu_0H$ at $T = 2$ K, respectively. (d) SdH oscillations as extracted from the $\rho_{xx}$ of a second sample as a function of $(\mu_0H)^{-1}$ at $T = 2$ K. Bottom inset: Fourier transform of the oscillatory signal. Top inset: $\rho_{xx}$ as a function of $(\mu_0H)^{2}$. Red-dashed line is a linear fit. (e) $\rho_{xx}$ in a logarithmic scale as a function of $T$ under various magnetic fields applied perpendicularly to the electrical current.
$\rho_{xx}$ shows a minimum at $T^{*}$ as indicated by the red arrow. Inset: $T^{*}$ as a function of $\mu_0H$ indicating a linear dependence in field.}
\end{center}
\end{figure*}
\begin{figure}[htb]
\begin{center}
\includegraphics[width= 7 cm]{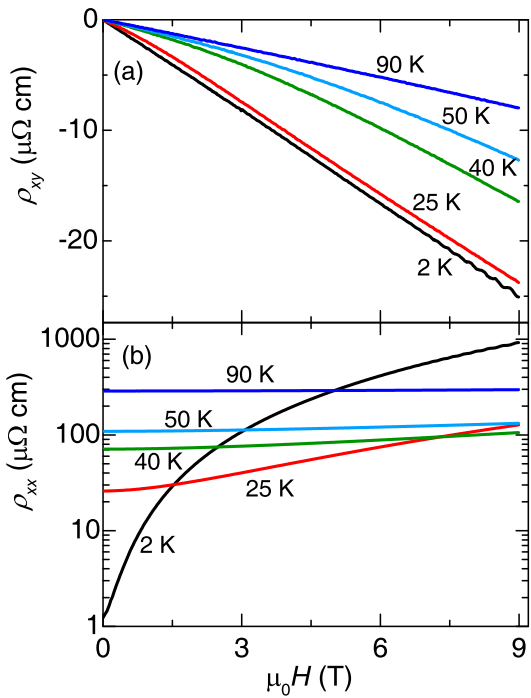}
\caption{(a) Hall resistivity of a $\gamma-$MoTe$_2$ single-crystal as a function of the field $\mu_0H$ and for several temperatures ranging from $T=2$ K to 90 K. A negative Hall
resistivity indicates that electrons dominate the transport at all temperatures. (b) Longitudinal resistivity $\rho_{xx}$ as a function of $\mu_0H$ and for the same temperatures. }
\end{center}
\end{figure}
\begin{figure}[htb]
\begin{center}
\includegraphics[width= 6.5 cm]{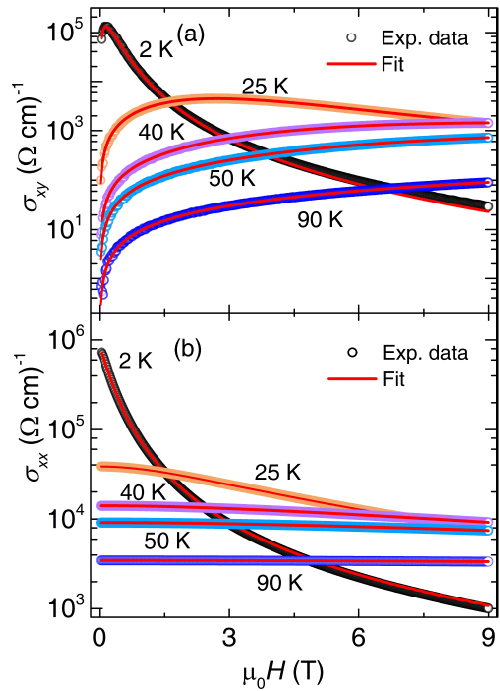}
\caption{Components of the conductivity tensor, i.e $\sigma_{xy}$ and $\sigma_{xx}$ in panels (a) and (b) respectively, as functions of the the magnetic field for temperatures ranging from 2 to 90 K. Open circles represent experimental data and red solid lines the fitting curves based on the two-carrier model. }
\end{center}
\end{figure}
\begin{figure*}[htb]
\begin{center}
\includegraphics[width= 14 cm]{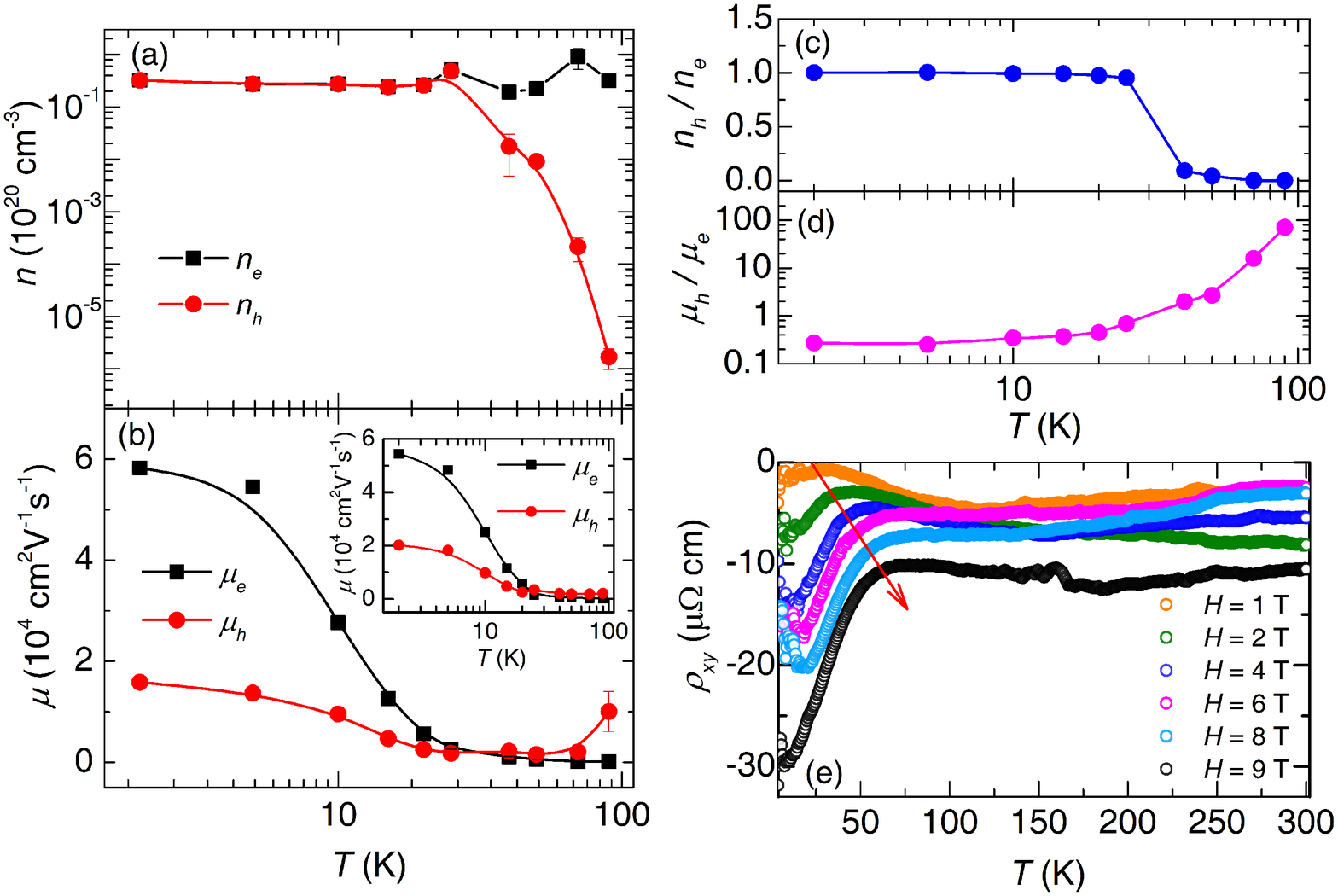}
\caption{(a) Density of electrons $n_e$ and density of holes $n_h$ extracted
from the two-carrier model analysis of $\sigma_{xy}$.
(b) Carrier mobility $\mu_e$ and $\mu_h$ as a function of temperature
deducted from $\sigma_{xy}$ (main panel) and from $\sigma_{xx}$(inset). (c) Density ratio and (d) mobility ratio between holes and
electrons as a function of $T$. (e) Hall resistivity as
a function of $T$ at various fields. A charge decrease in $\sigma_{xy}$ is indicated by the red
arrow. }
\end{center}
\end{figure*}
Therefore, understanding the electronic structure of $\gamma-$MoTe$_2$ is a critical step towards clarifying its topological nature, the properties of its superconducting state, and the mechanism leading to its enormous magnetoresistivity. Here, we investigated the electronic structure of $\gamma-$MoTe$_2$ through
a systematic temperature dependent study of the Hall-effect in high-quality $\gamma-$MoTe$_2$ single-crystals. Our goal is to determine fundamental electrical transport variables such as the density of carriers and their mobilities as well as their evolution as a function of the temperature.  Knowledge about these variables is crucial for understanding the electronic structure of this compound and for clarifying whether near perfect charge compensation, which is claimed to lead to the enormous magnetoresistivity \cite{ali2014large} of WTe$_2$, is pertinent also to $\gamma-$MoTe$_2$.

Single-crystals of MoTe$_2$ were grown at temperatures above 900 $^{\circ}$C \emph{via} a flux method using excess Te \cite{rhodes2016impurity}.
The magnetic field dependence of the longitudinal resistivity $\rho_{xx}$ and of the hall resistivity $\rho_{xy}= V_Ht/I_{xx}$, where $V_H$, $t$ and $I_{xx}$ are the Hall voltage, sample thickness, and electrical current respectively, were measured under magnetic fields up to $H=9$ T and under temperatures ranging from $T = 2$ K to $100$ K. Measurements were performed in a Quantum Design Physical Properties Measurement System using a conventional four-terminal configuration when measuring the resistivity, or a six-wire one when studying the Hall-effect. The electrical current was applied along the crystallographic $a-$axis. Any mixture of $\rho_{xx}$ and $\rho_{xy}$ was corrected by reversing the direction of the applied magnetic field: $\rho_{xy}$ is determined by subtracting the negative field trace from the positive field one.
A careful fitting of the Hall conductivity $\sigma_{xy} = - \rho_{xy}/(\rho_{xy}^2+\rho_{xx}^2) $ and of the longitudinal conductivity $\sigma_{xx} = \rho_{xx}/(\rho_{xy}^2+\rho_{xx}^2)$ based on the two-carrier model \cite{ali2014large}, yields the densities and mobilities of both electrons and holes as a function of the temperature. As we discuss below, we observe an excellent agreement between our experimental data and the two-carrier model particularly at low temperatures confirming the coexistence of electrons and holes in $\gamma-$MoTe$_2$. A sharp change in carrier densities is observed at temperatures between 30 and 70 K indicating a possible electronic phase-transition. The density of holes increases dramatically below $\sim 40$ K, reaching parity with the electron density below $\sim 15$ K. Therefore, near perfect charge-carrier compensation is an important ingredient for the colossal magnetoresistivity observed in $\gamma-$MoTe$_2$ at low temperatures.

Figure 1 (a) displays $\rho_{xx}$ measured under zero magnetic field as a function of the temperature, with $T$ ranging from $2$ to
300 K. For this particular crystal one obtains a residual resistivity ratio $RRR \equiv \rho$(300K)/$\rho$(2K)
= 1064, which is nearly one order of magnitude larger than previously reported values \cite{hughes1978electrical,keum2015bandgap,qi2015superconductivity,zandt2007quadratic}. The observed hysteresis around 250 K between the warm-up and the cool-down curves is associated with the structural phase-transition from the monoclinic
1$T'$ to the orthorhombic $T_{d}$ structure. The upper inset displays a fit of $\rho_{xx}(T)$ to a combination of Fermi liquid and electron-phonon scattering mechanisms, or a fit to $\rho_{xx}(T)=\rho_{0}+ aT^{2}+ bT^{5}$  where $\rho_{0} = \rho$($T = 0$ K) with $a$ and $b$ being fitting parameters. Figures 1 (b) and 1 (c) display the magnetic field dependence of both $\rho_{xx}$ and $\rho_{xy}$, respectively. $\rho_{xx}$ follows a nearly quadratic dependence on magnetic field
showing no signs of saturation. The resulting MR ratio $\Delta$MR $\equiv\left[\rho_{xx}(9\text{ T})-\rho_{xx}(0 \text{ T})\right]/\rho_{xx}(0 \text{ T})$ reaches 74791 \% at 2 K, which is the highest value reported so far for $\gamma-$MoTe$_2$ under $\mu_0H = 9$ T \cite{hughes1978electrical,keum2015bandgap,qi2015superconductivity}. In Ref. \onlinecite{rhodes2016impurity} we include measurements up to $H = 65 $ T in a lower quality crystal, which reveals a non-saturating $\Delta$MR $> 10^6$ \%.
As seen in Figs. 1(b) and 1(c), Shubnikov de Haas (SdH) oscillations are observed in both $\rho_{xx}(\mu_0H)$ and $\rho_{xy}(\mu_0H)$, respectively.
Figure 1(d) shows the SdH oscillations superimposed onto the $\rho_{xx}$ trace taken at $T=2$ K from a sample
with a $RRR$ of approximately 1900 and a $\Delta$MR $\sim$ 265340 \% at 9 T. $\rho_{xx}$ follows a near $(\mu_0H)^{2}$ dependence as seen in the inset.
Two frequencies $\alpha$ = 231 T and $\beta$ = 242 T are extracted from the Fast Fourier Transform of the oscillatory signal (bottom inset). A detailed analysis of the SdH effect can be found in Ref. \onlinecite{rhodes2016impurity}.
The temperature dependence of $\rho_{xx}$ under several magnetic fields is shown in Fig. 1(e). The observed behavior is similar to the one displayed by WTe$_2$.
Namely, under a magnetic field  $\rho_{xx}(T)$ essentially follows the zero-field curve until it reaches a minimum at a field dependent temperature $T^{*}$.
Below $T^{*}$, $\rho_{xx}$ increases rapidly upon cooling. With increasing fields $T^{*}$ shifts to higher temperatures at a rate of 2.01 K/T.

Figure 2 displays $\rho_{xy}$ and $\rho_{xx}$ as functions of $\mu_0 H$ for selected temperatures ranging from $T=2$ K to 90 K. Notice
that $\rho_{xy}$ remains nearly unchanged as the temperature increases from 2 to 15 K. Its negative sign indicates that the
electronic transport is dominated by electrons, or that the electrons have higher mobilities than the holes.
Below $\sim 20$ K, as well as above 70 K, $\rho_{xy}$ follows a linear dependence on $\mu_0H$ for fields up to $\mu_0H=9$ T.
Between 20 and 70 K, the Hall resistivity shows non-linear behavior, particularly at low fields, which is a clear indication for electrical conduction by both types of carriers.
$\rho_{xx}$ on the other hand, shows a quadratic field dependence below $T \sim 45$ K which becomes linear at higher temperatures.

Subsequently, we analyze the Hall response through the isotropic two-carrier model which was successfully used to describe
the contributions of holes and electrons to the electrical transport properties of
a number of compounds\cite{rullier2009hall,rullier2010hole,luo2015hall,takahashi2011low,xia2013indications}.
In this model, the conductivity tensor, in its complex representation, is given by:\cite{ali2014large}
\begin{equation}
\boldsymbol{\sigma} = e\left[\frac{n_e\mu_e}{1+i\mu_e \mu_0H} + \frac{n_h\mu_h}{1-i\mu_h \mu_0H}\right],\label{eq:First}
\end{equation}
Here, $n_e$ (or $n_h$) and $\mu_e$ (or $\mu_h$) are the densities of electrons (or holes) and the mobilities of electrons (or holes), respectively.
To appropriately evaluate the carrier densities and their mobilities, we obtained the Hall conductivity $\sigma_{xy}$
and the longitudinal conductivity $\sigma_{xx}$ from the original experimental data, as previously defined, and as shown in Figs. 3 (a) and 3 (b).

In the next step we fit $\sigma_{xy}$ and $\sigma_{xx}$ to the two-carrier
model, where the field dependence of the conductivity tensor is
given by \cite{huang2015observationTaAs,takahashi2011low,xia2013indications}
\begin{equation}
\sigma_{xy} = \left[n_e\mu_e^{2}\frac{1}{1 + (\mu_e \mu_0H)^{2}}-n_{h}\mu_h^{2}\frac{1}{1 + (\mu_h \mu_0H)^{2}}\right]e\mu_0H ,\label{eq:Third}
\end{equation}
\begin{equation}
\sigma_{xx} = \frac{n_e e\mu_e}{1 + (\mu_e \mu_0H)^{2}} + \frac{n_h e\mu_h}{1 + (\mu_h \mu_0H)^{2}} .\label{eq:Second}
\end{equation}

The results of the fittings of $\sigma_{xy}$ and $\sigma_{xx}$ to Eqs.(2) and (3) are displayed in Figs. 3 (a) and 3 (b), respectively.
The parameters resulting from our fittings can be found in the Supplemental Material \cite{Supplemental}. The excellent
agreement between our experimental data and the fittings to the two-carrier model, over a broad range of temperatures (see Fig. 3), confirms the coexistence of electrons
and holes in $\gamma-$MoTe$_2$.

The densities and mobilities of electrons and holes, as extracted from the fits of $\sigma_{xy}$ (as well as $\sigma_{xx}$) to the two-carrier model, are displayed in Figs. 4 (a) and 4 (b). The parameters extracted from $\sigma_{xy}$ and $\sigma_{xx}$ are close in value over a broad range of temperatures, particularly in what concerns the mobility, as illustrated by Fig. 4 (b) and its inset. In Fig. 4 (a) notice the drastic increase in $n_h$ as $T$ is reduced below $\sim 40$ K while $n_e$ remains nearly constant. At low-temperatures, the densities of electrons and holes become comparable while the mobility of the electrons becomes slightly higher than that of the holes.
An equal concentration of electrons and holes is consistent with the observation of a linear Hall resistivity below $\sim 20$ K.
Figure 4 (b) presents electron and hole mobilities as extracted from either $\sigma_{xy}$ or $\sigma_{xx}$ (inset). Both $\mu_e$ and $\mu_h$ increase dramatically, i.e.
by one order of magnitude, as $T$ decreases below $\sim 40$ K.  At $T = 2$ K, the extracted mobilities are $\mu_e \sim 5.8 \times10^{4}$ cm$^{2}$/Vs
and $\mu_h \sim 1.6 \times10^{4}$ cm$^{2}$/Vs. At $T= 2$ K the Hall mobility, or the ratio between the Hall constant as extracted from a linear fit of the Hall-effect and the resistivity, is $\mu_H \sim 2.3 \times 10^{4}$ cm$^{2}$/Vs, which is comparable to the values extracted from the two-carrier model, see Supplemental Material\cite{Supplemental}. The extracted ratio between carrier densities $n_h/n_e$, and the ratio between carrier mobilities, or $\mu_h/\mu_e$ are displayed in Figs. 4 (c) and 4 (d), respectively.
$n_h/n_e \simeq 1$ for temperatures below $T \simeq 40$ K but it decreases quickly as $T$ increases beyond this value.
The mobility ratio indicates that the hole-mobility considerably exceeds the electron one at high temperatures.
Finally, Fig. 4 (e) shows the Hall resistivity $\rho_{xy}$ as a function of $T$ under magnetic fields ranging from $\mu_0H = 1$ to 9 T. Below a field-dependent temperature $T^{\text{neg}}$,  $\rho_{xy}$ shows a pronounced decrement towards negative values. It turns out that below $T^{\text{neg}}$ one observes a sharp increase in the mobilities of both carriers and in the density ratio $n_h/n_e$. These observations, coupled to an anomaly seen in the heat capacity \cite{rhodes2016impurity} around these temperatures, suggest a possible electronic phase-transition analogous to the $T$-dependent Lifshitz transition \cite{wu2015temperature} reported for WTe$_2$.

In summary, our results indicate the coexistence of electrons and holes in $\gamma-$MoTe$_2$, with the density of holes increasing dramatically below $\sim 40$ K and finally reaching parity with the electron one below $\sim 15$ K. An equal density of electrons and holes is consistent with the observation of a linear Hall-resistivity below $T=20$ K which implies that $\gamma-$MoTe$_2$ is in fact better compensated than WTe$_2$ \cite{luo2015hall,rhodes2015role} whose Hall resistivity becomes non-linear at low temperatures. Hence, the extremely large and non-saturating magnetoresistance seen in $\gamma-$MoTe$_2$ should be primarily ascribed to the nearly perfect compensation between the densities of electrons and holes. The analysis of the Hall-effect through the two-carrier model yields high electron- and hole-mobilities, that is exceeding 10$^{4}$cm$^{2}$/Vs at $T=2$ K. Anomalies observed in the carrier densities, carrier mobilities, and in the heat capacity as a function of the temperature, suggest an electronic crossover or perhaps a phase-transition occurring between $\sim 30$ K and $\sim 70$ K, which is likely to lead to the electron-hole compensation and hence to the extremely large magnetoresistivity seen in $\gamma-$MoTe$_2$. It is important to clarify the existence and the role of this crossover, or phase-transition, since it is likely to affect the electronic structure at the Fermi level of semi-metallic MoTe$_2$ and its predicted topological properties \cite{soluyanov2015type,chang2016prediction}.

This work was supported by DOE-BES (for high field measurements) and
the U.S. Army Research Office through the MURI Grant W911NF-11-10362
(for synthesis and characterization). The NHMFL is supported by NSF
through NSF-DMR-1157490 and the State of Florida.

\bibliography{APL}

\begin{thebibliography}{39}%
\makeatletter
\providecommand \@ifxundefined [1]{%
 \@ifx{#1\undefined}
}%
\providecommand \@ifnum [1]{%
 \ifnum #1\expandafter \@firstoftwo
 \else \expandafter \@secondoftwo
 \fi
}%
\providecommand \@ifx [1]{%
 \ifx #1\expandafter \@firstoftwo
 \else \expandafter \@secondoftwo
 \fi
}%
\providecommand \natexlab [1]{#1}%
\providecommand \enquote  [1]{``#1''}%
\providecommand \bibnamefont  [1]{#1}%
\providecommand \bibfnamefont [1]{#1}%
\providecommand \citenamefont [1]{#1}%
\providecommand \href@noop [0]{\@secondoftwo}%
\providecommand \href [0]{\begingroup \@sanitize@url \@href}%
\providecommand \@href[1]{\@@startlink{#1}\@@href}%
\providecommand \@@href[1]{\endgroup#1\@@endlink}%
\providecommand \@sanitize@url [0]{\catcode `\\12\catcode `\$12\catcode
  `\&12\catcode `\#12\catcode `\^12\catcode `\_12\catcode `\%12\relax}%
\providecommand \@@startlink[1]{}%
\providecommand \@@endlink[0]{}%
\providecommand \url  [0]{\begingroup\@sanitize@url \@url }%
\providecommand \@url [1]{\endgroup\@href {#1}{\urlprefix }}%
\providecommand \urlprefix  [0]{URL }%
\providecommand \Eprint [0]{\href }%
\providecommand \doibase [0]{http://dx.doi.org/}%
\providecommand \selectlanguage [0]{\@gobble}%
\providecommand \bibinfo  [0]{\@secondoftwo}%
\providecommand \bibfield  [0]{\@secondoftwo}%
\providecommand \translation [1]{[#1]}%
\providecommand \BibitemOpen [0]{}%
\providecommand \bibitemStop [0]{}%
\providecommand \bibitemNoStop [0]{.\EOS\space}%
\providecommand \EOS [0]{\spacefactor3000\relax}%
\providecommand \BibitemShut  [1]{\csname bibitem#1\endcsname}%
\let\auto@bib@innerbib\@empty
\bibitem [{\citenamefont {Huang}\ \emph
  {et~al.}(2015{\natexlab{a}})\citenamefont {Huang}, \citenamefont {Xu},
  \citenamefont {Belopolski}, \citenamefont {Lee}, \citenamefont {Chang},
  \citenamefont {Wang}, \citenamefont {Alidoust}, \citenamefont {Bian},
  \citenamefont {Neupane}, \citenamefont {Zhang}, \citenamefont {Jia},
  \citenamefont {Bansil}, \citenamefont {Lin},\ and\ \citenamefont
  {Hasan}}]{huang2015weyl}%
  \BibitemOpen
  \bibfield  {author} {\bibinfo {author} {\bibfnamefont {S.-M.}\ \bibnamefont
  {Huang}}, \bibinfo {author} {\bibfnamefont {S.-Y.}\ \bibnamefont {Xu}},
  \bibinfo {author} {\bibfnamefont {I.}~\bibnamefont {Belopolski}}, \bibinfo
  {author} {\bibfnamefont {C.-C.}\ \bibnamefont {Lee}}, \bibinfo {author}
  {\bibfnamefont {G.}~\bibnamefont {Chang}}, \bibinfo {author} {\bibfnamefont
  {B.}~\bibnamefont {Wang}}, \bibinfo {author} {\bibfnamefont {N.}~\bibnamefont
  {Alidoust}}, \bibinfo {author} {\bibfnamefont {G.}~\bibnamefont {Bian}},
  \bibinfo {author} {\bibfnamefont {M.}~\bibnamefont {Neupane}}, \bibinfo
  {author} {\bibfnamefont {C.}~\bibnamefont {Zhang}}, \bibinfo {author}
  {\bibfnamefont {S.}~\bibnamefont {Jia}}, \bibinfo {author} {\bibfnamefont
  {A.}~\bibnamefont {Bansil}}, \bibinfo {author} {\bibfnamefont
  {H.}~\bibnamefont {Lin}}, \ and\ \bibinfo {author} {\bibfnamefont {M.~Z.}\
  \bibnamefont {Hasan}},\ }\href@noop {} {\bibfield  {journal} {\bibinfo
  {journal} {Nat. Commun.}\ }\textbf {\bibinfo {volume} {6}} (\bibinfo {year}
  {2015}{\natexlab{a}})}\BibitemShut {NoStop}%
\bibitem [{\citenamefont {Lv}\ \emph {et~al.}(2015)\citenamefont {Lv},
  \citenamefont {Weng}, \citenamefont {Fu}, \citenamefont {Wang}, \citenamefont
  {Miao}, \citenamefont {Ma}, \citenamefont {Richard}, \citenamefont {Huang},
  \citenamefont {Zhao}, \citenamefont {Chen}, \citenamefont {Fang},
  \citenamefont {Dai}, \citenamefont {Qian},\ and\ \citenamefont
  {Ding}}]{lv2015experimental}%
  \BibitemOpen
  \bibfield  {author} {\bibinfo {author} {\bibfnamefont {B.}~\bibnamefont
  {Lv}}, \bibinfo {author} {\bibfnamefont {H.}~\bibnamefont {Weng}}, \bibinfo
  {author} {\bibfnamefont {B.}~\bibnamefont {Fu}}, \bibinfo {author}
  {\bibfnamefont {X.}~\bibnamefont {Wang}}, \bibinfo {author} {\bibfnamefont
  {H.}~\bibnamefont {Miao}}, \bibinfo {author} {\bibfnamefont {J.}~\bibnamefont
  {Ma}}, \bibinfo {author} {\bibfnamefont {P.}~\bibnamefont {Richard}},
  \bibinfo {author} {\bibfnamefont {X.}~\bibnamefont {Huang}}, \bibinfo
  {author} {\bibfnamefont {L.}~\bibnamefont {Zhao}}, \bibinfo {author}
  {\bibfnamefont {G.}~\bibnamefont {Chen}}, \bibinfo {author} {\bibfnamefont
  {Z.}~\bibnamefont {Fang}}, \bibinfo {author} {\bibfnamefont {X.}~\bibnamefont
  {Dai}}, \bibinfo {author} {\bibfnamefont {T.}~\bibnamefont {Qian}}, \ and\
  \bibinfo {author} {\bibfnamefont {H.}~\bibnamefont {Ding}},\ }\href@noop {}
  {\bibfield  {journal} {\bibinfo  {journal} {Phys. Rev. X}\ }\textbf {\bibinfo
  {volume} {5}},\ \bibinfo {pages} {031013} (\bibinfo {year}
  {2015})}\BibitemShut {NoStop}%
\bibitem [{\citenamefont {Huang}\ \emph
  {et~al.}(2015{\natexlab{b}})\citenamefont {Huang}, \citenamefont {Zhao},
  \citenamefont {Long}, \citenamefont {Wang}, \citenamefont {Chen},
  \citenamefont {Yang}, \citenamefont {Liang}, \citenamefont {Xue},
  \citenamefont {Weng}, \citenamefont {Fang}, \citenamefont {Dai},\ and\
  \citenamefont {Chen}}]{huang2015observationTaAs}%
  \BibitemOpen
  \bibfield  {author} {\bibinfo {author} {\bibfnamefont {X.}~\bibnamefont
  {Huang}}, \bibinfo {author} {\bibfnamefont {L.}~\bibnamefont {Zhao}},
  \bibinfo {author} {\bibfnamefont {Y.}~\bibnamefont {Long}}, \bibinfo {author}
  {\bibfnamefont {P.}~\bibnamefont {Wang}}, \bibinfo {author} {\bibfnamefont
  {D.}~\bibnamefont {Chen}}, \bibinfo {author} {\bibfnamefont {Z.}~\bibnamefont
  {Yang}}, \bibinfo {author} {\bibfnamefont {H.}~\bibnamefont {Liang}},
  \bibinfo {author} {\bibfnamefont {M.}~\bibnamefont {Xue}}, \bibinfo {author}
  {\bibfnamefont {H.}~\bibnamefont {Weng}}, \bibinfo {author} {\bibfnamefont
  {Z.}~\bibnamefont {Fang}}, \bibinfo {author} {\bibfnamefont {X.}~\bibnamefont
  {Dai}}, \ and\ \bibinfo {author} {\bibfnamefont {G.}~\bibnamefont {Chen}},\
  }\href@noop {} {\bibfield  {journal} {\bibinfo  {journal} {Phys. Rev. X}\
  }\textbf {\bibinfo {volume} {5}},\ \bibinfo {pages} {031023} (\bibinfo {year}
  {2015}{\natexlab{b}})}\BibitemShut {NoStop}%
\bibitem [{\citenamefont {Xu}\ \emph {et~al.}(2015{\natexlab{a}})\citenamefont
  {Xu}, \citenamefont {Alidoust}, \citenamefont {Belopolski}, \citenamefont
  {Yuan}, \citenamefont {Bian}, \citenamefont {Chang}, \citenamefont {Zheng},
  \citenamefont {Strocov}, \citenamefont {Sanchez}, \citenamefont {Chang},
  \citenamefont {Zhang}, \citenamefont {Mou}, \citenamefont {Wu}, \citenamefont
  {Huang}, \citenamefont {Lee}, \citenamefont {Huang}, \citenamefont {Wang},
  \citenamefont {Bansil}, \citenamefont {Jeng}, \citenamefont {Neupert},
  \citenamefont {Kaminski}, \citenamefont {Lin}, \citenamefont {Jia},\ and\
  \citenamefont {Zahid~Hasan}}]{xu2015discoveryNbAs}%
  \BibitemOpen
  \bibfield  {author} {\bibinfo {author} {\bibfnamefont {S.-Y.}\ \bibnamefont
  {Xu}}, \bibinfo {author} {\bibfnamefont {N.}~\bibnamefont {Alidoust}},
  \bibinfo {author} {\bibfnamefont {I.}~\bibnamefont {Belopolski}}, \bibinfo
  {author} {\bibfnamefont {Z.}~\bibnamefont {Yuan}}, \bibinfo {author}
  {\bibfnamefont {G.}~\bibnamefont {Bian}}, \bibinfo {author} {\bibfnamefont
  {T.-R.}\ \bibnamefont {Chang}}, \bibinfo {author} {\bibfnamefont
  {H.}~\bibnamefont {Zheng}}, \bibinfo {author} {\bibfnamefont {V.~N.}\
  \bibnamefont {Strocov}}, \bibinfo {author} {\bibfnamefont {D.~S.}\
  \bibnamefont {Sanchez}}, \bibinfo {author} {\bibfnamefont {G.}~\bibnamefont
  {Chang}}, \bibinfo {author} {\bibfnamefont {C.}~\bibnamefont {Zhang}},
  \bibinfo {author} {\bibfnamefont {D.}~\bibnamefont {Mou}}, \bibinfo {author}
  {\bibfnamefont {Y.}~\bibnamefont {Wu}}, \bibinfo {author} {\bibfnamefont
  {L.}~\bibnamefont {Huang}}, \bibinfo {author} {\bibfnamefont {C.-C.}\
  \bibnamefont {Lee}}, \bibinfo {author} {\bibfnamefont {S.-M.}\ \bibnamefont
  {Huang}}, \bibinfo {author} {\bibfnamefont {B.}~\bibnamefont {Wang}},
  \bibinfo {author} {\bibfnamefont {A.}~\bibnamefont {Bansil}}, \bibinfo
  {author} {\bibfnamefont {H.-T.}\ \bibnamefont {Jeng}}, \bibinfo {author}
  {\bibfnamefont {T.}~\bibnamefont {Neupert}}, \bibinfo {author} {\bibfnamefont
  {A.}~\bibnamefont {Kaminski}}, \bibinfo {author} {\bibfnamefont
  {H.}~\bibnamefont {Lin}}, \bibinfo {author} {\bibfnamefont {S.}~\bibnamefont
  {Jia}}, \ and\ \bibinfo {author} {\bibfnamefont {M.}~\bibnamefont
  {Zahid~Hasan}},\ }\href@noop {} {\bibfield  {journal} {\bibinfo  {journal}
  {Nat. Phys.}\ }\textbf {\bibinfo {volume} {11}} (\bibinfo {year}
  {2015}{\natexlab{a}})}\BibitemShut {NoStop}%
\bibitem [{\citenamefont {Xu}\ \emph {et~al.}(2015{\natexlab{b}})\citenamefont
  {Xu}, \citenamefont {Belopolski}, \citenamefont {Alidoust}, \citenamefont
  {Neupane}, \citenamefont {Bian}, \citenamefont {Zhang}, \citenamefont
  {Sankar}, \citenamefont {Chang}, \citenamefont {Yuan}, \citenamefont {Lee},
  \citenamefont {Huang}, \citenamefont {Zheng}, \citenamefont {Ma},
  \citenamefont {Sanchez}, \citenamefont {Wang}, \citenamefont {Bansil},
  \citenamefont {Chou}, \citenamefont {Shibayev}, \citenamefont {Lin},
  \citenamefont {Jia},\ and\ \citenamefont {Hasan}}]{xu2015discoverywely}%
  \BibitemOpen
  \bibfield  {author} {\bibinfo {author} {\bibfnamefont {S.-Y.}\ \bibnamefont
  {Xu}}, \bibinfo {author} {\bibfnamefont {I.}~\bibnamefont {Belopolski}},
  \bibinfo {author} {\bibfnamefont {N.}~\bibnamefont {Alidoust}}, \bibinfo
  {author} {\bibfnamefont {M.}~\bibnamefont {Neupane}}, \bibinfo {author}
  {\bibfnamefont {G.}~\bibnamefont {Bian}}, \bibinfo {author} {\bibfnamefont
  {C.}~\bibnamefont {Zhang}}, \bibinfo {author} {\bibfnamefont
  {R.}~\bibnamefont {Sankar}}, \bibinfo {author} {\bibfnamefont
  {G.}~\bibnamefont {Chang}}, \bibinfo {author} {\bibfnamefont
  {Z.}~\bibnamefont {Yuan}}, \bibinfo {author} {\bibfnamefont {C.-C.}\
  \bibnamefont {Lee}}, \bibinfo {author} {\bibfnamefont {S.-M.}\ \bibnamefont
  {Huang}}, \bibinfo {author} {\bibfnamefont {H.}~\bibnamefont {Zheng}},
  \bibinfo {author} {\bibfnamefont {J.}~\bibnamefont {Ma}}, \bibinfo {author}
  {\bibfnamefont {D.~S.}\ \bibnamefont {Sanchez}}, \bibinfo {author}
  {\bibfnamefont {B.}~\bibnamefont {Wang}}, \bibinfo {author} {\bibfnamefont
  {A.}~\bibnamefont {Bansil}}, \bibinfo {author} {\bibfnamefont
  {F.}~\bibnamefont {Chou}}, \bibinfo {author} {\bibfnamefont {P.~P.}\
  \bibnamefont {Shibayev}}, \bibinfo {author} {\bibfnamefont {H.}~\bibnamefont
  {Lin}}, \bibinfo {author} {\bibfnamefont {S.}~\bibnamefont {Jia}}, \ and\
  \bibinfo {author} {\bibfnamefont {M.~Z.}\ \bibnamefont {Hasan}},\ }\href@noop
  {} {\bibfield  {journal} {\bibinfo  {journal} {Science}\ }\textbf {\bibinfo
  {volume} {349}},\ \bibinfo {pages} {613} (\bibinfo {year}
  {2015}{\natexlab{b}})}\BibitemShut {NoStop}%
\bibitem [{\citenamefont {Jin}\ \emph {et~al.}(1994)\citenamefont {Jin},
  \citenamefont {McCormack}, \citenamefont {Tiefel},\ and\ \citenamefont
  {Ramesh}}]{jin1994colossal}%
  \BibitemOpen
  \bibfield  {author} {\bibinfo {author} {\bibfnamefont {S.}~\bibnamefont
  {Jin}}, \bibinfo {author} {\bibfnamefont {M.}~\bibnamefont {McCormack}},
  \bibinfo {author} {\bibfnamefont {T.}~\bibnamefont {Tiefel}}, \ and\ \bibinfo
  {author} {\bibfnamefont {R.}~\bibnamefont {Ramesh}},\ }\href@noop {}
  {\bibfield  {journal} {\bibinfo  {journal} {J. Appl. Phys.}\ }\textbf
  {\bibinfo {volume} {76}},\ \bibinfo {pages} {6929} (\bibinfo {year}
  {1994})}\BibitemShut {NoStop}%
\bibitem [{\citenamefont {Ramirez}\ \emph {et~al.}(1997)\citenamefont
  {Ramirez}, \citenamefont {Cava},\ and\ \citenamefont
  {Krajewski}}]{ramirez1997colossal}%
  \BibitemOpen
  \bibfield  {author} {\bibinfo {author} {\bibfnamefont {A.}~\bibnamefont
  {Ramirez}}, \bibinfo {author} {\bibfnamefont {R.}~\bibnamefont {Cava}}, \
  and\ \bibinfo {author} {\bibfnamefont {J.}~\bibnamefont {Krajewski}},\
  }\href@noop {} {\bibfield  {journal} {\bibinfo  {journal} {Nature}\ }\textbf
  {\bibinfo {volume} {386}},\ \bibinfo {pages} {156} (\bibinfo {year}
  {1997})}\BibitemShut {NoStop}%
\bibitem [{\citenamefont {Matsumoto}\ \emph {et~al.}(2001)\citenamefont
  {Matsumoto}, \citenamefont {Murakami}, \citenamefont {Shono}, \citenamefont
  {Hasegawa}, \citenamefont {Fukumura}, \citenamefont {Kawasaki}, \citenamefont
  {Ahmet}, \citenamefont {Chikyow}, \citenamefont {Koshihara},\ and\
  \citenamefont {Koinuma}}]{matsumoto2001room}%
  \BibitemOpen
  \bibfield  {author} {\bibinfo {author} {\bibfnamefont {Y.}~\bibnamefont
  {Matsumoto}}, \bibinfo {author} {\bibfnamefont {M.}~\bibnamefont {Murakami}},
  \bibinfo {author} {\bibfnamefont {T.}~\bibnamefont {Shono}}, \bibinfo
  {author} {\bibfnamefont {T.}~\bibnamefont {Hasegawa}}, \bibinfo {author}
  {\bibfnamefont {T.}~\bibnamefont {Fukumura}}, \bibinfo {author}
  {\bibfnamefont {M.}~\bibnamefont {Kawasaki}}, \bibinfo {author}
  {\bibfnamefont {P.}~\bibnamefont {Ahmet}}, \bibinfo {author} {\bibfnamefont
  {T.}~\bibnamefont {Chikyow}}, \bibinfo {author} {\bibfnamefont {S.-y.}\
  \bibnamefont {Koshihara}}, \ and\ \bibinfo {author} {\bibfnamefont
  {H.}~\bibnamefont {Koinuma}},\ }\href@noop {} {\bibfield  {journal} {\bibinfo
   {journal} {Science}\ }\textbf {\bibinfo {volume} {291}},\ \bibinfo {pages}
  {854} (\bibinfo {year} {2001})}\BibitemShut {NoStop}%
\bibitem [{\citenamefont {Prinz}(1998)}]{prinz1998magnetoelectronics}%
  \BibitemOpen
  \bibfield  {author} {\bibinfo {author} {\bibfnamefont {G.~A.}\ \bibnamefont
  {Prinz}},\ }\href@noop {} {\bibfield  {journal} {\bibinfo  {journal}
  {Science}\ }\textbf {\bibinfo {volume} {282}},\ \bibinfo {pages} {1660}
  (\bibinfo {year} {1998})}\BibitemShut {NoStop}%
\bibitem [{\citenamefont {Parkin}\ \emph {et~al.}(2003)\citenamefont {Parkin},
  \citenamefont {Jiang}, \citenamefont {Kaiser}, \citenamefont {Panchula},
  \citenamefont {Roche},\ and\ \citenamefont
  {Samant}}]{parkin2003magnetically}%
  \BibitemOpen
  \bibfield  {author} {\bibinfo {author} {\bibfnamefont {S.}~\bibnamefont
  {Parkin}}, \bibinfo {author} {\bibfnamefont {X.}~\bibnamefont {Jiang}},
  \bibinfo {author} {\bibfnamefont {C.}~\bibnamefont {Kaiser}}, \bibinfo
  {author} {\bibfnamefont {A.}~\bibnamefont {Panchula}}, \bibinfo {author}
  {\bibfnamefont {K.}~\bibnamefont {Roche}}, \ and\ \bibinfo {author}
  {\bibfnamefont {M.}~\bibnamefont {Samant}},\ }\href@noop {} {\bibfield
  {journal} {\bibinfo  {journal} {Proc. IEEE}\ }\textbf {\bibinfo {volume}
  {91}},\ \bibinfo {pages} {661} (\bibinfo {year} {2003})}\BibitemShut
  {NoStop}%
\bibitem [{\citenamefont {Soluyanov}\ \emph {et~al.}(2015)\citenamefont
  {Soluyanov}, \citenamefont {Gresch}, \citenamefont {Wang}, \citenamefont
  {Wu}, \citenamefont {Troyer}, \citenamefont {Dai},\ and\ \citenamefont
  {Bernevig}}]{soluyanov2015type}%
  \BibitemOpen
  \bibfield  {author} {\bibinfo {author} {\bibfnamefont {A.~A.}\ \bibnamefont
  {Soluyanov}}, \bibinfo {author} {\bibfnamefont {D.}~\bibnamefont {Gresch}},
  \bibinfo {author} {\bibfnamefont {Z.}~\bibnamefont {Wang}}, \bibinfo {author}
  {\bibfnamefont {Q.}~\bibnamefont {Wu}}, \bibinfo {author} {\bibfnamefont
  {M.}~\bibnamefont {Troyer}}, \bibinfo {author} {\bibfnamefont
  {X.}~\bibnamefont {Dai}}, \ and\ \bibinfo {author} {\bibfnamefont {B.~A.}\
  \bibnamefont {Bernevig}},\ }\href@noop {} {\bibfield  {journal} {\bibinfo
  {journal} {Nature}\ }\textbf {\bibinfo {volume} {527}},\ \bibinfo {pages}
  {495} (\bibinfo {year} {2015})}\BibitemShut {NoStop}%
\bibitem [{\citenamefont {Chang}\ \emph {et~al.}(2016)\citenamefont {Chang},
  \citenamefont {Xu}, \citenamefont {Chang}, \citenamefont {Lee}, \citenamefont
  {Huang}, \citenamefont {Wang}, \citenamefont {Bian}, \citenamefont {Zheng},
  \citenamefont {Sanchez}, \citenamefont {Belopolski}, \citenamefont
  {Alidoust}, \citenamefont {Neupane}, \citenamefont {Bansil}, \citenamefont
  {Jeng}, \citenamefont {Lin},\ and\ \citenamefont
  {Zahid~Hasan}}]{chang2016prediction}%
  \BibitemOpen
  \bibfield  {author} {\bibinfo {author} {\bibfnamefont {T.-R.}\ \bibnamefont
  {Chang}}, \bibinfo {author} {\bibfnamefont {S.-Y.}\ \bibnamefont {Xu}},
  \bibinfo {author} {\bibfnamefont {G.}~\bibnamefont {Chang}}, \bibinfo
  {author} {\bibfnamefont {C.-C.}\ \bibnamefont {Lee}}, \bibinfo {author}
  {\bibfnamefont {S.-M.}\ \bibnamefont {Huang}}, \bibinfo {author}
  {\bibfnamefont {B.}~\bibnamefont {Wang}}, \bibinfo {author} {\bibfnamefont
  {G.}~\bibnamefont {Bian}}, \bibinfo {author} {\bibfnamefont {H.}~\bibnamefont
  {Zheng}}, \bibinfo {author} {\bibfnamefont {D.~S.}\ \bibnamefont {Sanchez}},
  \bibinfo {author} {\bibfnamefont {I.}~\bibnamefont {Belopolski}}, \bibinfo
  {author} {\bibfnamefont {N.}~\bibnamefont {Alidoust}}, \bibinfo {author}
  {\bibfnamefont {M.}~\bibnamefont {Neupane}}, \bibinfo {author} {\bibfnamefont
  {A.}~\bibnamefont {Bansil}}, \bibinfo {author} {\bibfnamefont {H.-T.}\
  \bibnamefont {Jeng}}, \bibinfo {author} {\bibfnamefont {H.}~\bibnamefont
  {Lin}}, \ and\ \bibinfo {author} {\bibfnamefont {M.}~\bibnamefont
  {Zahid~Hasan}},\ }\href@noop {} {\bibfield  {journal} {\bibinfo  {journal}
  {Nat. Commun.}\ }\textbf {\bibinfo {volume} {7}} (\bibinfo {year}
  {2016})}\BibitemShut {NoStop}%
\bibitem [{\citenamefont {Ali}\ \emph {et~al.}(2014)\citenamefont {Ali},
  \citenamefont {Xiong}, \citenamefont {Flynn}, \citenamefont {Tao},
  \citenamefont {Gibson}, \citenamefont {Schoop}, \citenamefont {Liang},
  \citenamefont {Haldolaarachchige}, \citenamefont {Hirschberger},
  \citenamefont {Ong},\ and\ \citenamefont {Cava}}]{ali2014large}%
  \BibitemOpen
  \bibfield  {author} {\bibinfo {author} {\bibfnamefont {M.~N.}\ \bibnamefont
  {Ali}}, \bibinfo {author} {\bibfnamefont {J.}~\bibnamefont {Xiong}}, \bibinfo
  {author} {\bibfnamefont {S.}~\bibnamefont {Flynn}}, \bibinfo {author}
  {\bibfnamefont {J.}~\bibnamefont {Tao}}, \bibinfo {author} {\bibfnamefont
  {Q.~D.}\ \bibnamefont {Gibson}}, \bibinfo {author} {\bibfnamefont {L.~M.}\
  \bibnamefont {Schoop}}, \bibinfo {author} {\bibfnamefont {T.}~\bibnamefont
  {Liang}}, \bibinfo {author} {\bibfnamefont {N.}~\bibnamefont
  {Haldolaarachchige}}, \bibinfo {author} {\bibfnamefont {M.}~\bibnamefont
  {Hirschberger}}, \bibinfo {author} {\bibfnamefont {N.~P.}\ \bibnamefont
  {Ong}}, \ and\ \bibinfo {author} {\bibfnamefont {R.~J.}\ \bibnamefont
  {Cava}},\ }\href@noop {} {\bibfield  {journal} {\bibinfo  {journal} {Nature}\
  }\textbf {\bibinfo {volume} {514}},\ \bibinfo {pages} {205} (\bibinfo {year}
  {2014})}\BibitemShut {NoStop}%
\bibitem [{\citenamefont {Keum}\ \emph {et~al.}(2015)\citenamefont {Keum},
  \citenamefont {Cho}, \citenamefont {Kim}, \citenamefont {Choe}, \citenamefont
  {Sung}, \citenamefont {Kan}, \citenamefont {Kang}, \citenamefont {Hwang},
  \citenamefont {Kim}, \citenamefont {Yang}, \citenamefont {Chang},\ and\
  \citenamefont {Lee}}]{keum2015bandgap}%
  \BibitemOpen
  \bibfield  {author} {\bibinfo {author} {\bibfnamefont {D.~H.}\ \bibnamefont
  {Keum}}, \bibinfo {author} {\bibfnamefont {S.}~\bibnamefont {Cho}}, \bibinfo
  {author} {\bibfnamefont {J.~H.}\ \bibnamefont {Kim}}, \bibinfo {author}
  {\bibfnamefont {D.-H.}\ \bibnamefont {Choe}}, \bibinfo {author}
  {\bibfnamefont {H.-J.}\ \bibnamefont {Sung}}, \bibinfo {author}
  {\bibfnamefont {M.}~\bibnamefont {Kan}}, \bibinfo {author} {\bibfnamefont
  {H.}~\bibnamefont {Kang}}, \bibinfo {author} {\bibfnamefont {J.-Y.}\
  \bibnamefont {Hwang}}, \bibinfo {author} {\bibfnamefont {S.~W.}\ \bibnamefont
  {Kim}}, \bibinfo {author} {\bibfnamefont {H.}~\bibnamefont {Yang}}, \bibinfo
  {author} {\bibfnamefont {K.~J.}\ \bibnamefont {Chang}}, \ and\ \bibinfo
  {author} {\bibfnamefont {Y.~H.}\ \bibnamefont {Lee}},\ }\href@noop {}
  {\bibfield  {journal} {\bibinfo  {journal} {Nat. Phys.}\ }\textbf {\bibinfo
  {volume} {11}},\ \bibinfo {pages} {482} (\bibinfo {year} {2015})}\BibitemShut
  {NoStop}%
\bibitem [{\citenamefont {Rhodes}\ \emph {et~al.}(2015)\citenamefont {Rhodes},
  \citenamefont {Das}, \citenamefont {Zhang}, \citenamefont {Zeng},
  \citenamefont {Pradhan}, \citenamefont {Kikugawa}, \citenamefont
  {Manousakis},\ and\ \citenamefont {Balicas}}]{rhodes2015role}%
  \BibitemOpen
  \bibfield  {author} {\bibinfo {author} {\bibfnamefont {D.}~\bibnamefont
  {Rhodes}}, \bibinfo {author} {\bibfnamefont {S.}~\bibnamefont {Das}},
  \bibinfo {author} {\bibfnamefont {Q.}~\bibnamefont {Zhang}}, \bibinfo
  {author} {\bibfnamefont {B.}~\bibnamefont {Zeng}}, \bibinfo {author}
  {\bibfnamefont {N.}~\bibnamefont {Pradhan}}, \bibinfo {author} {\bibfnamefont
  {N.}~\bibnamefont {Kikugawa}}, \bibinfo {author} {\bibfnamefont
  {E.}~\bibnamefont {Manousakis}}, \ and\ \bibinfo {author} {\bibfnamefont
  {L.}~\bibnamefont {Balicas}},\ }\href@noop {} {\bibfield  {journal} {\bibinfo
   {journal} {Phys. Rev. B}\ }\textbf {\bibinfo {volume} {92}},\ \bibinfo
  {pages} {125152} (\bibinfo {year} {2015})}\BibitemShut {NoStop}%
\bibitem [{\citenamefont {Hughes}\ and\ \citenamefont
  {Friend}(1978)}]{hughes1978electrical}%
  \BibitemOpen
  \bibfield  {author} {\bibinfo {author} {\bibfnamefont {H.}~\bibnamefont
  {Hughes}}\ and\ \bibinfo {author} {\bibfnamefont {R.}~\bibnamefont
  {Friend}},\ }\href@noop {} {\bibfield  {journal} {\bibinfo  {journal} {J.
  Phys. C Solid State}\ }\textbf {\bibinfo {volume} {11}},\ \bibinfo {pages}
  {L103} (\bibinfo {year} {1978})}\BibitemShut {NoStop}%
\bibitem [{\citenamefont {Canadell}\ and\ \citenamefont
  {Whangbo}(1990)}]{canadell1990semimetallic}%
  \BibitemOpen
  \bibfield  {author} {\bibinfo {author} {\bibfnamefont {E.}~\bibnamefont
  {Canadell}}\ and\ \bibinfo {author} {\bibfnamefont {M.~H.}\ \bibnamefont
  {Whangbo}},\ }\href@noop {} {\bibfield  {journal} {\bibinfo  {journal}
  {Inorg. Chem.}\ }\textbf {\bibinfo {volume} {29}},\ \bibinfo {pages} {1398}
  (\bibinfo {year} {1990})}\BibitemShut {NoStop}%
\bibitem [{\citenamefont {Zandt}\ \emph {et~al.}(2007)\citenamefont {Zandt},
  \citenamefont {Dwelk}, \citenamefont {Janowitz},\ and\ \citenamefont
  {Manzke}}]{zandt2007quadratic}%
  \BibitemOpen
  \bibfield  {author} {\bibinfo {author} {\bibfnamefont {T.}~\bibnamefont
  {Zandt}}, \bibinfo {author} {\bibfnamefont {H.}~\bibnamefont {Dwelk}},
  \bibinfo {author} {\bibfnamefont {C.}~\bibnamefont {Janowitz}}, \ and\
  \bibinfo {author} {\bibfnamefont {R.}~\bibnamefont {Manzke}},\ }\href@noop {}
  {\bibfield  {journal} {\bibinfo  {journal} {J. Alloy Compd.}\ }\textbf
  {\bibinfo {volume} {442}},\ \bibinfo {pages} {216} (\bibinfo {year}
  {2007})}\BibitemShut {NoStop}%
\bibitem [{\citenamefont {Rhodes}\ \emph {et~al.}(2016)\citenamefont {Rhodes},
  \citenamefont {Zhou}, \citenamefont {Schönemann}, \citenamefont {Zhang},
  \citenamefont {Kampert}, \citenamefont {Shimura}, \citenamefont {McCandless},
  \citenamefont {Chan}, \citenamefont {Das}, \citenamefont {Manousakis},
  \citenamefont {Johannes},\ and\ \citenamefont
  {Balicas}}]{rhodes2016impurity}%
  \BibitemOpen
  \bibfield  {author} {\bibinfo {author} {\bibfnamefont {D.}~\bibnamefont
  {Rhodes}}, \bibinfo {author} {\bibfnamefont {Q.}~\bibnamefont {Zhou}},
  \bibinfo {author} {\bibfnamefont {R.}~\bibnamefont {Schönemann}}, \bibinfo
  {author} {\bibfnamefont {Q.~R.}\ \bibnamefont {Zhang}}, \bibinfo {author}
  {\bibfnamefont {E.}~\bibnamefont {Kampert}}, \bibinfo {author} {\bibfnamefont
  {Y.}~\bibnamefont {Shimura}}, \bibinfo {author} {\bibfnamefont {G.~T.}\
  \bibnamefont {McCandless}}, \bibinfo {author} {\bibfnamefont {J.~Y.}\
  \bibnamefont {Chan}}, \bibinfo {author} {\bibfnamefont {S.}~\bibnamefont
  {Das}}, \bibinfo {author} {\bibfnamefont {E.}~\bibnamefont {Manousakis}},
  \bibinfo {author} {\bibfnamefont {M.~D.}\ \bibnamefont {Johannes}}, \ and\
  \bibinfo {author} {\bibfnamefont {L.}~\bibnamefont {Balicas}},\ }\href@noop
  {} {\bibfield  {journal} {\bibinfo  {journal} {arXiv preprint
  arXiv:1605.09065}\ } (\bibinfo {year} {2016})}\BibitemShut {NoStop}%
\bibitem [{\citenamefont {Pletikosi{\'c}}\ \emph {et~al.}(2014)\citenamefont
  {Pletikosi{\'c}}, \citenamefont {Ali}, \citenamefont {Fedorov}, \citenamefont
  {Cava},\ and\ \citenamefont {Valla}}]{pletikosic2014electronic}%
  \BibitemOpen
  \bibfield  {author} {\bibinfo {author} {\bibfnamefont {I.}~\bibnamefont
  {Pletikosi{\'c}}}, \bibinfo {author} {\bibfnamefont {M.~N.}\ \bibnamefont
  {Ali}}, \bibinfo {author} {\bibfnamefont {A.}~\bibnamefont {Fedorov}},
  \bibinfo {author} {\bibfnamefont {R.}~\bibnamefont {Cava}}, \ and\ \bibinfo
  {author} {\bibfnamefont {T.}~\bibnamefont {Valla}},\ }\href@noop {}
  {\bibfield  {journal} {\bibinfo  {journal} {Phys. Rev. Lett.}\ }\textbf
  {\bibinfo {volume} {113}},\ \bibinfo {pages} {216601} (\bibinfo {year}
  {2014})}\BibitemShut {NoStop}%
\bibitem [{\citenamefont {Jiang}\ \emph {et~al.}(2015)\citenamefont {Jiang},
  \citenamefont {Tang}, \citenamefont {Pan}, \citenamefont {Liu}, \citenamefont
  {Niu}, \citenamefont {Wang}, \citenamefont {Xu}, \citenamefont {Yang},
  \citenamefont {Xie}, \citenamefont {Song}, \citenamefont {Dudin},
  \citenamefont {Kim}, \citenamefont {Hoesch}, \citenamefont {Das},
  \citenamefont {Vobornik}, \citenamefont {Wan},\ and\ \citenamefont
  {Feng}}]{jiang2015signature}%
  \BibitemOpen
  \bibfield  {author} {\bibinfo {author} {\bibfnamefont {J.}~\bibnamefont
  {Jiang}}, \bibinfo {author} {\bibfnamefont {F.}~\bibnamefont {Tang}},
  \bibinfo {author} {\bibfnamefont {X.}~\bibnamefont {Pan}}, \bibinfo {author}
  {\bibfnamefont {H.}~\bibnamefont {Liu}}, \bibinfo {author} {\bibfnamefont
  {X.}~\bibnamefont {Niu}}, \bibinfo {author} {\bibfnamefont {Y.}~\bibnamefont
  {Wang}}, \bibinfo {author} {\bibfnamefont {D.}~\bibnamefont {Xu}}, \bibinfo
  {author} {\bibfnamefont {H.}~\bibnamefont {Yang}}, \bibinfo {author}
  {\bibfnamefont {B.}~\bibnamefont {Xie}}, \bibinfo {author} {\bibfnamefont
  {F.}~\bibnamefont {Song}}, \bibinfo {author} {\bibfnamefont {P.}~\bibnamefont
  {Dudin}}, \bibinfo {author} {\bibfnamefont {T.}~\bibnamefont {Kim}}, \bibinfo
  {author} {\bibfnamefont {M.}~\bibnamefont {Hoesch}}, \bibinfo {author}
  {\bibfnamefont {P.~K.}\ \bibnamefont {Das}}, \bibinfo {author} {\bibfnamefont
  {I.}~\bibnamefont {Vobornik}}, \bibinfo {author} {\bibfnamefont
  {X.}~\bibnamefont {Wan}}, \ and\ \bibinfo {author} {\bibfnamefont
  {D.}~\bibnamefont {Feng}},\ }\href@noop {} {\bibfield  {journal} {\bibinfo
  {journal} {Phys. Rev. Lett.}\ }\textbf {\bibinfo {volume} {115}},\ \bibinfo
  {pages} {166601} (\bibinfo {year} {2015})}\BibitemShut {NoStop}%
\bibitem [{\citenamefont {Cai}\ \emph {et~al.}(2015)\citenamefont {Cai},
  \citenamefont {Hu}, \citenamefont {He}, \citenamefont {Pan}, \citenamefont
  {Hong}, \citenamefont {Zhang}, \citenamefont {Zhang}, \citenamefont {Wei},
  \citenamefont {Mao},\ and\ \citenamefont {Li}}]{cai2015drastic}%
  \BibitemOpen
  \bibfield  {author} {\bibinfo {author} {\bibfnamefont {P.}~\bibnamefont
  {Cai}}, \bibinfo {author} {\bibfnamefont {J.}~\bibnamefont {Hu}}, \bibinfo
  {author} {\bibfnamefont {L.}~\bibnamefont {He}}, \bibinfo {author}
  {\bibfnamefont {J.}~\bibnamefont {Pan}}, \bibinfo {author} {\bibfnamefont
  {X.}~\bibnamefont {Hong}}, \bibinfo {author} {\bibfnamefont {Z.}~\bibnamefont
  {Zhang}}, \bibinfo {author} {\bibfnamefont {J.}~\bibnamefont {Zhang}},
  \bibinfo {author} {\bibfnamefont {J.}~\bibnamefont {Wei}}, \bibinfo {author}
  {\bibfnamefont {Z.}~\bibnamefont {Mao}}, \ and\ \bibinfo {author}
  {\bibfnamefont {S.}~\bibnamefont {Li}},\ }\href@noop {} {\bibfield  {journal}
  {\bibinfo  {journal} {Phys. Rev. Lett.}\ }\textbf {\bibinfo {volume} {115}},\
  \bibinfo {pages} {057202} (\bibinfo {year} {2015})}\BibitemShut {NoStop}%
\bibitem [{\citenamefont {Luo}\ \emph {et~al.}(2015)\citenamefont {Luo},
  \citenamefont {Li}, \citenamefont {Dai}, \citenamefont {Miao}, \citenamefont
  {Shi}, \citenamefont {Ding}, \citenamefont {Taylor}, \citenamefont
  {Yarotski}, \citenamefont {Prasankumar},\ and\ \citenamefont
  {Thompson}}]{luo2015hall}%
  \BibitemOpen
  \bibfield  {author} {\bibinfo {author} {\bibfnamefont {Y.}~\bibnamefont
  {Luo}}, \bibinfo {author} {\bibfnamefont {H.}~\bibnamefont {Li}}, \bibinfo
  {author} {\bibfnamefont {Y.}~\bibnamefont {Dai}}, \bibinfo {author}
  {\bibfnamefont {H.}~\bibnamefont {Miao}}, \bibinfo {author} {\bibfnamefont
  {Y.}~\bibnamefont {Shi}}, \bibinfo {author} {\bibfnamefont {H.}~\bibnamefont
  {Ding}}, \bibinfo {author} {\bibfnamefont {A.}~\bibnamefont {Taylor}},
  \bibinfo {author} {\bibfnamefont {D.}~\bibnamefont {Yarotski}}, \bibinfo
  {author} {\bibfnamefont {R.}~\bibnamefont {Prasankumar}}, \ and\ \bibinfo
  {author} {\bibfnamefont {J.}~\bibnamefont {Thompson}},\ }\href@noop {}
  {\bibfield  {journal} {\bibinfo  {journal} {Appl Phys Lett}\ }\textbf
  {\bibinfo {volume} {107}},\ \bibinfo {pages} {182411} (\bibinfo {year}
  {2015})}\BibitemShut {NoStop}%
\bibitem [{\citenamefont {Revolinsky}\ and\ \citenamefont
  {Beerntsen}(1966)}]{revolinsky1966electrical}%
  \BibitemOpen
  \bibfield  {author} {\bibinfo {author} {\bibfnamefont {E.}~\bibnamefont
  {Revolinsky}}\ and\ \bibinfo {author} {\bibfnamefont {D.}~\bibnamefont
  {Beerntsen}},\ }\href@noop {} {\bibfield  {journal} {\bibinfo  {journal} {J.
  Phys. Chem. Solids}\ }\textbf {\bibinfo {volume} {27}},\ \bibinfo {pages}
  {523} (\bibinfo {year} {1966})}\BibitemShut {NoStop}%
\bibitem [{\citenamefont {Dawson}\ and\ \citenamefont
  {Bullett}(1987)}]{dawson1987electronic}%
  \BibitemOpen
  \bibfield  {author} {\bibinfo {author} {\bibfnamefont {W.}~\bibnamefont
  {Dawson}}\ and\ \bibinfo {author} {\bibfnamefont {D.}~\bibnamefont
  {Bullett}},\ }\href@noop {} {\bibfield  {journal} {\bibinfo  {journal} {J.
  Phys. C Solid State}\ }\textbf {\bibinfo {volume} {20}},\ \bibinfo {pages}
  {6159} (\bibinfo {year} {1987})}\BibitemShut {NoStop}%
\bibitem [{\citenamefont {Clarke}\ \emph {et~al.}(1978)\citenamefont {Clarke},
  \citenamefont {Marseglia},\ and\ \citenamefont {Hughes}}]{clarke1978low}%
  \BibitemOpen
  \bibfield  {author} {\bibinfo {author} {\bibfnamefont {R.}~\bibnamefont
  {Clarke}}, \bibinfo {author} {\bibfnamefont {E.}~\bibnamefont {Marseglia}}, \
  and\ \bibinfo {author} {\bibfnamefont {H.}~\bibnamefont {Hughes}},\
  }\href@noop {} {\bibfield  {journal} {\bibinfo  {journal} {Philos. Mag. B}\
  }\textbf {\bibinfo {volume} {38}},\ \bibinfo {pages} {121} (\bibinfo {year}
  {1978})}\BibitemShut {NoStop}%
\bibitem [{\citenamefont {Qi}\ \emph {et~al.}(2015)\citenamefont {Qi},
  \citenamefont {Naumov}, \citenamefont {Ali}, \citenamefont {Rajamathi},
  \citenamefont {Schnelle}, \citenamefont {Barkalov}, \citenamefont {Hanfland},
  \citenamefont {Wu}, \citenamefont {Shekhar}, \citenamefont {Sun},
  \citenamefont {Süß}, \citenamefont {Schmidt}, \citenamefont {Schwarz},
  \citenamefont {Pippel}, \citenamefont {Werner}, \citenamefont {Hillebrand},
  \citenamefont {Förster}, \citenamefont {Kampert}, \citenamefont {Parkin},
  \citenamefont {Cava}, \citenamefont {Felser}, \citenamefont {Yan},\ and\
  \citenamefont {Medvedev}}]{qi2015superconductivity}%
  \BibitemOpen
  \bibfield  {author} {\bibinfo {author} {\bibfnamefont {Y.}~\bibnamefont
  {Qi}}, \bibinfo {author} {\bibfnamefont {P.~G.}\ \bibnamefont {Naumov}},
  \bibinfo {author} {\bibfnamefont {M.~N.}\ \bibnamefont {Ali}}, \bibinfo
  {author} {\bibfnamefont {C.~R.}\ \bibnamefont {Rajamathi}}, \bibinfo {author}
  {\bibfnamefont {W.}~\bibnamefont {Schnelle}}, \bibinfo {author}
  {\bibfnamefont {O.}~\bibnamefont {Barkalov}}, \bibinfo {author}
  {\bibfnamefont {M.}~\bibnamefont {Hanfland}}, \bibinfo {author}
  {\bibfnamefont {S.-C.}\ \bibnamefont {Wu}}, \bibinfo {author} {\bibfnamefont
  {C.}~\bibnamefont {Shekhar}}, \bibinfo {author} {\bibfnamefont
  {Y.}~\bibnamefont {Sun}}, \bibinfo {author} {\bibfnamefont {V.}~\bibnamefont
  {Süß}}, \bibinfo {author} {\bibfnamefont {M.}~\bibnamefont {Schmidt}},
  \bibinfo {author} {\bibfnamefont {U.}~\bibnamefont {Schwarz}}, \bibinfo
  {author} {\bibfnamefont {E.}~\bibnamefont {Pippel}}, \bibinfo {author}
  {\bibfnamefont {P.}~\bibnamefont {Werner}}, \bibinfo {author} {\bibfnamefont
  {R.}~\bibnamefont {Hillebrand}}, \bibinfo {author} {\bibfnamefont
  {T.}~\bibnamefont {Förster}}, \bibinfo {author} {\bibfnamefont
  {E.}~\bibnamefont {Kampert}}, \bibinfo {author} {\bibfnamefont
  {S.}~\bibnamefont {Parkin}}, \bibinfo {author} {\bibfnamefont {R.~J.}\
  \bibnamefont {Cava}}, \bibinfo {author} {\bibfnamefont {C.}~\bibnamefont
  {Felser}}, \bibinfo {author} {\bibfnamefont {B.}~\bibnamefont {Yan}}, \ and\
  \bibinfo {author} {\bibfnamefont {S.~A.}\ \bibnamefont {Medvedev}},\
  }\href@noop {} {\bibfield  {journal} {\bibinfo  {journal} {Nat.
  Commun.7,11038}\ } (\bibinfo {year} {2015})}\BibitemShut {NoStop}%
\bibitem [{\citenamefont {Sun}\ \emph {et~al.}(2015)\citenamefont {Sun},
  \citenamefont {Wu}, \citenamefont {Ali}, \citenamefont {Felser},\ and\
  \citenamefont {Yan}}]{sun2015prediction}%
  \BibitemOpen
  \bibfield  {author} {\bibinfo {author} {\bibfnamefont {Y.}~\bibnamefont
  {Sun}}, \bibinfo {author} {\bibfnamefont {S.-C.}\ \bibnamefont {Wu}},
  \bibinfo {author} {\bibfnamefont {M.~N.}\ \bibnamefont {Ali}}, \bibinfo
  {author} {\bibfnamefont {C.}~\bibnamefont {Felser}}, \ and\ \bibinfo {author}
  {\bibfnamefont {B.}~\bibnamefont {Yan}},\ }\href@noop {} {\bibfield
  {journal} {\bibinfo  {journal} {Phys. Rev. B}\ }\textbf {\bibinfo {volume}
  {92}},\ \bibinfo {pages} {161107} (\bibinfo {year} {2015})}\BibitemShut
  {NoStop}%
\bibitem [{\citenamefont {Xu}\ \emph {et~al.}(2016)\citenamefont {Xu},
  \citenamefont {Wang}, \citenamefont {Weber}, \citenamefont {Magrez},
  \citenamefont {Bugnon}, \citenamefont {Berger}, \citenamefont {Matt},
  \citenamefont {Ma}, \citenamefont {Fu}, \citenamefont {Lv}, \citenamefont
  {Plumb}, \citenamefont {Radovic}, \citenamefont {Pomjakushina}, \citenamefont
  {Conder}, \citenamefont {Qian}, \citenamefont {Dil}, \citenamefont {Mesot},
  \citenamefont {Ding},\ and\ \citenamefont {Shi}}]{xu2016discovery}%
  \BibitemOpen
  \bibfield  {author} {\bibinfo {author} {\bibfnamefont {N.}~\bibnamefont
  {Xu}}, \bibinfo {author} {\bibfnamefont {Z.~J.}\ \bibnamefont {Wang}},
  \bibinfo {author} {\bibfnamefont {A.~P.}\ \bibnamefont {Weber}}, \bibinfo
  {author} {\bibfnamefont {A.}~\bibnamefont {Magrez}}, \bibinfo {author}
  {\bibfnamefont {P.}~\bibnamefont {Bugnon}}, \bibinfo {author} {\bibfnamefont
  {H.}~\bibnamefont {Berger}}, \bibinfo {author} {\bibfnamefont {C.~E.}\
  \bibnamefont {Matt}}, \bibinfo {author} {\bibfnamefont {J.~Z.}\ \bibnamefont
  {Ma}}, \bibinfo {author} {\bibfnamefont {B.~B.}\ \bibnamefont {Fu}}, \bibinfo
  {author} {\bibfnamefont {B.~Q.}\ \bibnamefont {Lv}}, \bibinfo {author}
  {\bibfnamefont {N.~C.}\ \bibnamefont {Plumb}}, \bibinfo {author}
  {\bibfnamefont {M.}~\bibnamefont {Radovic}}, \bibinfo {author} {\bibfnamefont
  {E.}~\bibnamefont {Pomjakushina}}, \bibinfo {author} {\bibfnamefont
  {K.}~\bibnamefont {Conder}}, \bibinfo {author} {\bibfnamefont
  {T.}~\bibnamefont {Qian}}, \bibinfo {author} {\bibfnamefont {J.~H.}\
  \bibnamefont {Dil}}, \bibinfo {author} {\bibfnamefont {J.}~\bibnamefont
  {Mesot}}, \bibinfo {author} {\bibfnamefont {H.}~\bibnamefont {Ding}}, \ and\
  \bibinfo {author} {\bibfnamefont {M.}~\bibnamefont {Shi}},\ }\href@noop {}
  {\bibfield  {journal} {\bibinfo  {journal} {arXiv preprint arXiv:1604.02116}\
  } (\bibinfo {year} {2016})}\BibitemShut {NoStop}%
\bibitem [{\citenamefont {Liang}\ \emph {et~al.}(2016)\citenamefont {Liang},
  \citenamefont {Huang}, \citenamefont {Nie}, \citenamefont {Ding},
  \citenamefont {Gao}, \citenamefont {Hu}, \citenamefont {He}, \citenamefont
  {Zhang}, \citenamefont {Wang}, \citenamefont {Shen}, \citenamefont {Liu},
  \citenamefont {Ai}, \citenamefont {Yu}, \citenamefont {Sun}, \citenamefont
  {Zhao}, \citenamefont {Lv}, \citenamefont {Liu}, \citenamefont {Li},
  \citenamefont {Zhang}, \citenamefont {Hu}, \citenamefont {Xu}, \citenamefont
  {Zhao}, \citenamefont {Liu}, \citenamefont {Mao}, \citenamefont {Jia},
  \citenamefont {Zhang}, \citenamefont {Zhang}, \citenamefont {Yang},
  \citenamefont {Wang}, \citenamefont {Peng}, \citenamefont {Weng},
  \citenamefont {Dai}, \citenamefont {Fang}, \citenamefont {Xu}, \citenamefont
  {Chen},\ and\ \citenamefont {Zhou}}]{liang2016electronic}%
  \BibitemOpen
  \bibfield  {author} {\bibinfo {author} {\bibfnamefont {A.}~\bibnamefont
  {Liang}}, \bibinfo {author} {\bibfnamefont {J.}~\bibnamefont {Huang}},
  \bibinfo {author} {\bibfnamefont {S.}~\bibnamefont {Nie}}, \bibinfo {author}
  {\bibfnamefont {Y.}~\bibnamefont {Ding}}, \bibinfo {author} {\bibfnamefont
  {Q.}~\bibnamefont {Gao}}, \bibinfo {author} {\bibfnamefont {C.}~\bibnamefont
  {Hu}}, \bibinfo {author} {\bibfnamefont {S.}~\bibnamefont {He}}, \bibinfo
  {author} {\bibfnamefont {Y.}~\bibnamefont {Zhang}}, \bibinfo {author}
  {\bibfnamefont {C.}~\bibnamefont {Wang}}, \bibinfo {author} {\bibfnamefont
  {B.}~\bibnamefont {Shen}}, \bibinfo {author} {\bibfnamefont {J.}~\bibnamefont
  {Liu}}, \bibinfo {author} {\bibfnamefont {P.}~\bibnamefont {Ai}}, \bibinfo
  {author} {\bibfnamefont {L.}~\bibnamefont {Yu}}, \bibinfo {author}
  {\bibfnamefont {X.}~\bibnamefont {Sun}}, \bibinfo {author} {\bibfnamefont
  {W.}~\bibnamefont {Zhao}}, \bibinfo {author} {\bibfnamefont {S.}~\bibnamefont
  {Lv}}, \bibinfo {author} {\bibfnamefont {D.}~\bibnamefont {Liu}}, \bibinfo
  {author} {\bibfnamefont {C.}~\bibnamefont {Li}}, \bibinfo {author}
  {\bibfnamefont {Y.}~\bibnamefont {Zhang}}, \bibinfo {author} {\bibfnamefont
  {Y.}~\bibnamefont {Hu}}, \bibinfo {author} {\bibfnamefont {Y.}~\bibnamefont
  {Xu}}, \bibinfo {author} {\bibfnamefont {L.}~\bibnamefont {Zhao}}, \bibinfo
  {author} {\bibfnamefont {G.}~\bibnamefont {Liu}}, \bibinfo {author}
  {\bibfnamefont {Z.}~\bibnamefont {Mao}}, \bibinfo {author} {\bibfnamefont
  {X.}~\bibnamefont {Jia}}, \bibinfo {author} {\bibfnamefont {F.}~\bibnamefont
  {Zhang}}, \bibinfo {author} {\bibfnamefont {S.}~\bibnamefont {Zhang}},
  \bibinfo {author} {\bibfnamefont {F.}~\bibnamefont {Yang}}, \bibinfo {author}
  {\bibfnamefont {Z.}~\bibnamefont {Wang}}, \bibinfo {author} {\bibfnamefont
  {Q.}~\bibnamefont {Peng}}, \bibinfo {author} {\bibfnamefont {H.}~\bibnamefont
  {Weng}}, \bibinfo {author} {\bibfnamefont {X.}~\bibnamefont {Dai}}, \bibinfo
  {author} {\bibfnamefont {Z.}~\bibnamefont {Fang}}, \bibinfo {author}
  {\bibfnamefont {Z.}~\bibnamefont {Xu}}, \bibinfo {author} {\bibfnamefont
  {C.}~\bibnamefont {Chen}}, \ and\ \bibinfo {author} {\bibfnamefont {X.~J.}\
  \bibnamefont {Zhou}},\ }\href@noop {} {\bibfield  {journal} {\bibinfo
  {journal} {arXiv preprint arXiv:1604.01706}\ } (\bibinfo {year}
  {2016})}\BibitemShut {NoStop}%
\bibitem [{\citenamefont {Jiang}\ \emph {et~al.}(2016)\citenamefont {Jiang},
  \citenamefont {Liu}, \citenamefont {Sun}, \citenamefont {Yang}, \citenamefont
  {Rajamathi}, \citenamefont {Qi}, \citenamefont {Yang}, \citenamefont {Chen},
  \citenamefont {Peng}, \citenamefont {Hwang}, \citenamefont {Sun},
  \citenamefont {Mo}, \citenamefont {Vobornik}, \citenamefont {Fujii},
  \citenamefont {Parkin}, \citenamefont {Felser}, \citenamefont {Yan},\ and\
  \citenamefont {Chen}}]{jiang2016observation}%
  \BibitemOpen
  \bibfield  {author} {\bibinfo {author} {\bibfnamefont {J.}~\bibnamefont
  {Jiang}}, \bibinfo {author} {\bibfnamefont {Z.~K.}\ \bibnamefont {Liu}},
  \bibinfo {author} {\bibfnamefont {Y.}~\bibnamefont {Sun}}, \bibinfo {author}
  {\bibfnamefont {H.~F.}\ \bibnamefont {Yang}}, \bibinfo {author}
  {\bibfnamefont {R.}~\bibnamefont {Rajamathi}}, \bibinfo {author}
  {\bibfnamefont {Y.~P.}\ \bibnamefont {Qi}}, \bibinfo {author} {\bibfnamefont
  {L.~X.}\ \bibnamefont {Yang}}, \bibinfo {author} {\bibfnamefont
  {C.}~\bibnamefont {Chen}}, \bibinfo {author} {\bibfnamefont {H.}~\bibnamefont
  {Peng}}, \bibinfo {author} {\bibfnamefont {C.-C.}\ \bibnamefont {Hwang}},
  \bibinfo {author} {\bibfnamefont {S.~Z.}\ \bibnamefont {Sun}}, \bibinfo
  {author} {\bibfnamefont {S.-K.}\ \bibnamefont {Mo}}, \bibinfo {author}
  {\bibfnamefont {I.}~\bibnamefont {Vobornik}}, \bibinfo {author}
  {\bibfnamefont {J.}~\bibnamefont {Fujii}}, \bibinfo {author} {\bibfnamefont
  {S.~S.~P.}\ \bibnamefont {Parkin}}, \bibinfo {author} {\bibfnamefont
  {C.}~\bibnamefont {Felser}}, \bibinfo {author} {\bibfnamefont {B.~H.}\
  \bibnamefont {Yan}}, \ and\ \bibinfo {author} {\bibfnamefont {Y.~L.}\
  \bibnamefont {Chen}},\ }\href@noop {} {\bibfield  {journal} {\bibinfo
  {journal} {arXiv preprint arXiv:1604.00139}\ } (\bibinfo {year}
  {2016})}\BibitemShut {NoStop}%
\bibitem [{\citenamefont {Tamai}\ \emph {et~al.}(2016)\citenamefont {Tamai},
  \citenamefont {Wu}, \citenamefont {Cucchi}, \citenamefont {Bruno},
  \citenamefont {Ricco}, \citenamefont {Kim}, \citenamefont {Hoesch},
  \citenamefont {Barreteau}, \citenamefont {Giannini}, \citenamefont {Bernard},
  \citenamefont {Soluyanov},\ and\ \citenamefont
  {Baumberger}}]{tamai2016fermi}%
  \BibitemOpen
  \bibfield  {author} {\bibinfo {author} {\bibfnamefont {A.}~\bibnamefont
  {Tamai}}, \bibinfo {author} {\bibfnamefont {Q.~S.}\ \bibnamefont {Wu}},
  \bibinfo {author} {\bibfnamefont {I.}~\bibnamefont {Cucchi}}, \bibinfo
  {author} {\bibfnamefont {F.~Y.}\ \bibnamefont {Bruno}}, \bibinfo {author}
  {\bibfnamefont {S.}~\bibnamefont {Ricco}}, \bibinfo {author} {\bibfnamefont
  {T.~K.}\ \bibnamefont {Kim}}, \bibinfo {author} {\bibfnamefont
  {M.}~\bibnamefont {Hoesch}}, \bibinfo {author} {\bibfnamefont
  {C.}~\bibnamefont {Barreteau}}, \bibinfo {author} {\bibfnamefont
  {E.}~\bibnamefont {Giannini}}, \bibinfo {author} {\bibfnamefont
  {C.}~\bibnamefont {Bernard}}, \bibinfo {author} {\bibfnamefont {A.~A.}\
  \bibnamefont {Soluyanov}}, \ and\ \bibinfo {author} {\bibfnamefont
  {F.}~\bibnamefont {Baumberger}},\ }\href@noop {} {\bibfield  {journal}
  {\bibinfo  {journal} {arXiv preprint arXiv:1604.08228}\ } (\bibinfo {year}
  {2016})}\BibitemShut {NoStop}%
\bibitem [{\citenamefont {Huang}\ \emph {et~al.}(2016)\citenamefont {Huang},
  \citenamefont {{McCormick}}, \citenamefont {Ochi}, \citenamefont {Zhao},
  \citenamefont {Suzuki}, \citenamefont {Arita}, \citenamefont {Wu},
  \citenamefont {Mou}, \citenamefont {Cao}, \citenamefont {Yan}, \citenamefont
  {Trivedi},\ and\ \citenamefont {Kaminski}}]{huang2016spectroscopic}%
  \BibitemOpen
  \bibfield  {author} {\bibinfo {author} {\bibfnamefont {L.}~\bibnamefont
  {Huang}}, \bibinfo {author} {\bibfnamefont {T.~M.}\ \bibnamefont
  {{McCormick}}}, \bibinfo {author} {\bibfnamefont {M.}~\bibnamefont {Ochi}},
  \bibinfo {author} {\bibfnamefont {Z.}~\bibnamefont {Zhao}}, \bibinfo {author}
  {\bibfnamefont {M.-t.}\ \bibnamefont {Suzuki}}, \bibinfo {author}
  {\bibfnamefont {R.}~\bibnamefont {Arita}}, \bibinfo {author} {\bibfnamefont
  {Y.}~\bibnamefont {Wu}}, \bibinfo {author} {\bibfnamefont {D.}~\bibnamefont
  {Mou}}, \bibinfo {author} {\bibfnamefont {H.}~\bibnamefont {Cao}}, \bibinfo
  {author} {\bibfnamefont {J.}~\bibnamefont {Yan}}, \bibinfo {author}
  {\bibfnamefont {N.}~\bibnamefont {Trivedi}}, \ and\ \bibinfo {author}
  {\bibfnamefont {A.}~\bibnamefont {Kaminski}},\ }\href@noop {} {\bibfield
  {journal} {\bibinfo  {journal} {arXiv preprint arXiv:1603.06482}\ } (\bibinfo
  {year} {2016})}\BibitemShut {NoStop}%
\bibitem [{\citenamefont {Rullier-Albenque}\ \emph {et~al.}(2009)\citenamefont
  {Rullier-Albenque}, \citenamefont {Colson}, \citenamefont {Forget},\ and\
  \citenamefont {Alloul}}]{rullier2009hall}%
  \BibitemOpen
  \bibfield  {author} {\bibinfo {author} {\bibfnamefont {F.}~\bibnamefont
  {Rullier-Albenque}}, \bibinfo {author} {\bibfnamefont {D.}~\bibnamefont
  {Colson}}, \bibinfo {author} {\bibfnamefont {A.}~\bibnamefont {Forget}}, \
  and\ \bibinfo {author} {\bibfnamefont {H.}~\bibnamefont {Alloul}},\
  }\href@noop {} {\bibfield  {journal} {\bibinfo  {journal} {Phys. Rev. Lett.}\
  }\textbf {\bibinfo {volume} {103}},\ \bibinfo {pages} {057001} (\bibinfo
  {year} {2009})}\BibitemShut {NoStop}%
\bibitem [{\citenamefont {Rullier-Albenque}\ \emph {et~al.}(2010)\citenamefont
  {Rullier-Albenque}, \citenamefont {Colson}, \citenamefont {Forget},
  \citenamefont {Thu{\'e}ry},\ and\ \citenamefont
  {Poissonnet}}]{rullier2010hole}%
  \BibitemOpen
  \bibfield  {author} {\bibinfo {author} {\bibfnamefont {F.}~\bibnamefont
  {Rullier-Albenque}}, \bibinfo {author} {\bibfnamefont {D.}~\bibnamefont
  {Colson}}, \bibinfo {author} {\bibfnamefont {A.}~\bibnamefont {Forget}},
  \bibinfo {author} {\bibfnamefont {P.}~\bibnamefont {Thu{\'e}ry}}, \ and\
  \bibinfo {author} {\bibfnamefont {S.}~\bibnamefont {Poissonnet}},\
  }\href@noop {} {\bibfield  {journal} {\bibinfo  {journal} {Phys. Rev. B}\
  }\textbf {\bibinfo {volume} {81}},\ \bibinfo {pages} {224503} (\bibinfo
  {year} {2010})}\BibitemShut {NoStop}%
\bibitem [{\citenamefont {Takahashi}\ \emph {et~al.}(2011)\citenamefont
  {Takahashi}, \citenamefont {Okazaki}, \citenamefont {Yasui},\ and\
  \citenamefont {Terasaki}}]{takahashi2011low}%
  \BibitemOpen
  \bibfield  {author} {\bibinfo {author} {\bibfnamefont {H.}~\bibnamefont
  {Takahashi}}, \bibinfo {author} {\bibfnamefont {R.}~\bibnamefont {Okazaki}},
  \bibinfo {author} {\bibfnamefont {Y.}~\bibnamefont {Yasui}}, \ and\ \bibinfo
  {author} {\bibfnamefont {I.}~\bibnamefont {Terasaki}},\ }\href@noop {}
  {\bibfield  {journal} {\bibinfo  {journal} {Phys. Rev. B}\ }\textbf {\bibinfo
  {volume} {84}},\ \bibinfo {pages} {205215} (\bibinfo {year}
  {2011})}\BibitemShut {NoStop}%
\bibitem [{\citenamefont {Xia}\ \emph {et~al.}(2013)\citenamefont {Xia},
  \citenamefont {Ren}, \citenamefont {Sulaev}, \citenamefont {Liu},
  \citenamefont {Shen},\ and\ \citenamefont {Wang}}]{xia2013indications}%
  \BibitemOpen
  \bibfield  {author} {\bibinfo {author} {\bibfnamefont {B.}~\bibnamefont
  {Xia}}, \bibinfo {author} {\bibfnamefont {P.}~\bibnamefont {Ren}}, \bibinfo
  {author} {\bibfnamefont {A.}~\bibnamefont {Sulaev}}, \bibinfo {author}
  {\bibfnamefont {P.}~\bibnamefont {Liu}}, \bibinfo {author} {\bibfnamefont
  {S.-Q.}\ \bibnamefont {Shen}}, \ and\ \bibinfo {author} {\bibfnamefont
  {L.}~\bibnamefont {Wang}},\ }\href@noop {} {\bibfield  {journal} {\bibinfo
  {journal} {Phys. Rev. B}\ }\textbf {\bibinfo {volume} {87}},\ \bibinfo
  {pages} {085442} (\bibinfo {year} {2013})}\BibitemShut {NoStop}%
\bibitem [{Sup()}]{Supplemental}%
  \BibitemOpen
  \href@noop {} {\ }\bibinfo {note} {See Supplemental Material at
  http://link.aps.org/supplemental/ for the two-carrier model fitting
  parameters and Hall mobility data.}\BibitemShut {Stop}%
\bibitem [{\citenamefont {Wu}\ \emph {et~al.}(2015)\citenamefont {Wu},
  \citenamefont {Jo}, \citenamefont {Ochi}, \citenamefont {Huang},
  \citenamefont {Mou}, \citenamefont {Budko}, \citenamefont {Canfield},
  \citenamefont {Trivedi}, \citenamefont {Arita},\ and\ \citenamefont
  {Kaminski}}]{wu2015temperature}%
  \BibitemOpen
  \bibfield  {author} {\bibinfo {author} {\bibfnamefont {Y.}~\bibnamefont
  {Wu}}, \bibinfo {author} {\bibfnamefont {N.~H.}\ \bibnamefont {Jo}}, \bibinfo
  {author} {\bibfnamefont {M.}~\bibnamefont {Ochi}}, \bibinfo {author}
  {\bibfnamefont {L.}~\bibnamefont {Huang}}, \bibinfo {author} {\bibfnamefont
  {D.}~\bibnamefont {Mou}}, \bibinfo {author} {\bibfnamefont {S.~L.}\
  \bibnamefont {Budko}}, \bibinfo {author} {\bibfnamefont {P.}~\bibnamefont
  {Canfield}}, \bibinfo {author} {\bibfnamefont {N.}~\bibnamefont {Trivedi}},
  \bibinfo {author} {\bibfnamefont {R.}~\bibnamefont {Arita}}, \ and\ \bibinfo
  {author} {\bibfnamefont {A.}~\bibnamefont {Kaminski}},\ }\href@noop {}
  {\bibfield  {journal} {\bibinfo  {journal} {Phys. Rev. Lett.}\ }\textbf
  {\bibinfo {volume} {115}},\ \bibinfo {pages} {166602} (\bibinfo {year}
  {2015})}\BibitemShut {NoStop}%
\end{thebibliography}%
\end{document}